\documentclass[prd,preprint,showpacs,preprintnumbers,nofootinbib,eqsecnum,superscriptaddress]{revtex4}

 \usepackage[dvips,final]{graphicx}
  \usepackage{amssymb}
   \usepackage{amsmath}
    \usepackage{amsfonts}
     \usepackage{epsfig}
      \usepackage{bm}

\usepackage{mathpazo}

\usepackage[section]{placeins}

\usepackage{multirow}
\usepackage{ctable}
\usepackage{booktabs}
\usepackage{array}
\usepackage{tabularx}
\usepackage{xcolor}
\usepackage{pstricks}

\newcommand{\ket}[1]{ {#1} \rangle}
\newcommand{\bra}[1]{\langle {#1} }
\begin{document}

\title{Production of $\chi_c$ pairs in $k_T$-factorization}

\author{Anna Cisek}
\email{acisek@univ.rzeszow.pl}
\affiliation{Faculty of Mathematics and Natural Sciences,
University of Rzesz\'ow, ul. Pigonia 1, PL-35-310 Rzesz\'ow, Poland}

\author{Wolfgang Sch\"afer}
\email{Wolfgang.Schafer@ifj.edu.pl} \affiliation{Institute of Nuclear
Physics, Polish Academy of Sciences, ul. Radzikowskiego 152, PL-31-342 Krak{\'o}w, Poland}

\author{Antoni Szczurek\footnote{also at University of Rzesz\'ow, PL-35-959 Rzesz\'ow, Poland}}
\email{antoni.szczurek@ifj.edu.pl} 
\affiliation{Institute of Nuclear Physics, Polish Academy of Sciences, 
ul. Radzikowskiego 152, PL-31-342 Krak{\'o}w, Poland}

\begin{abstract}
We calculate the production of pairs of $\chi_c(J)$ mesons with
all possible combinations of $J=0,1,2$.
The leading order production mechanism is the crossed-channel
gluon exchange in the gluon-gluon fusion reaction.	

The building blocks are the vertices $g^* g^* \to \chi_c(J)$
for off shell gluons. We stick to the color-singlet model and 
calculate the gluon fusion vertices in the limit of heavy quarks
with nonrelativistic motion in the bound state.

These vertices are used to construct the 
$g^* g^* \to \chi_c(J_1) \chi_c(J_2)$ amplitudes.
We then calculate hadron-level cross sections using the $k_T$-factorization
approach. In our numerical predictions, we use the KMR-type unintegrated gluon distributions.
Several differential distributions
at the $pp$ center of mass energy $\sqrt{s} = 8 \, {\rm TeV}$ are shown.

The salient feature of the $t$ and $u$-channel gluon exchange are the 
broad distributions in rapidity difference $\Delta y$ between $\chi_c$ mesons.

\end{abstract}

\maketitle

\section{Introduction}
Recently, cross sections for the production of $J/\psi$-pairs were measured
at the Tevatron \cite{Abazov:2014qba} and 
the LHC \cite{CMS_jpsijpsi,ATLAS_jpsijpsi,Aaij:2016bqq,Aaij:2011yc}.
There remain a number of puzzles, especially with the CMS and ATLAS data.
Here the leading order of ${\cal{O}}(\alpha_S^4)$ (see e.g. \cite{JJ_kt,Baranov:2012re}) is clearly not 
sufficient. The double parton scattering (DPS) contribution was claimed to be large
or even dominant in some corners of the phase space, when the rapidity distance $\Delta y$ between
two $J/\psi$ mesons is large. However the effective cross sections $\sigma_{\rm eff}$ found 
from empirical analyses are about a factor $2.5$ smaller than the usually accepted 
$\sigma_{\rm eff} = 15 \, \rm{mb}$. It is an open issue at the moment whether this points to
a nonuniversality of $\sigma_{\rm eff}$ or whether there are additional single parton scattering
mechanisms which can alleviate the tension.

The production of quarkonium pairs is interesting in a broader context. Here we wish to
consider production of pairs of $\chi_c$ mesons. This process is more difficult to measure
experimentally but interesting from the theoretical point of view. 
A feed down to the double $J/\psi$ channel is interesting in the context of the puzzles 
mentioned above.

The single-inclusive $\chi_c$ meson production was a topic of both experimental \cite{Aaij:2013dja,Chatrchyan:2012ub,ATLAS:2014ala}
and theoretical \cite{Hagler:2000dd,Kniehl:2006sk,Likhoded:2014kfa,Baranov:2015yea,Cisek:2016uxz} studies. 
The cross section for single $\chi_c$ production is rather
large. The nonrelativistic perturbative QCD is the standard theoretical approach in this 
context. In leading order the gluon fusion $g^* g^* \to \chi_c(J), J=0,1,2$ is the underlying production
mechanism. The $k_T$-factorization approach provides a reasonable description of
the experimental data \cite{Baranov:2015yea,Cisek:2016uxz}.

In the present letter we shall include the production of all combinations of $\chi_c$ meson
pair production. The cross section will be calculated in $k_T$-factorization approach using
newly derived off-shell matrix elements for the $g^* g^* \to \chi_c(i) \chi_c(j)$ process.

A first evaluation of the total cross section will be given. We also show some differential 
distributions. 

\subsection{The $p p \to \chi_c(J_1) \chi_c(J_2) X$ reaction, formalism} 

\begin{figure}[!h]
\begin{minipage}{0.32\textwidth}
\centerline{\includegraphics[width=1.0\textwidth]{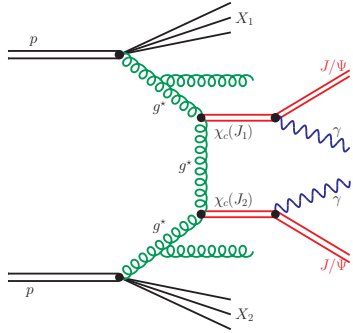}}
\end{minipage}
   \caption{A diagrammatic representation of the leading order 
mechanisms for $pp \to \chi_c(J_1) \chi_c(J_2) \to 
(J/\psi+\gamma) (J/\psi+\gamma)$ reaction.
 }
\label{fig:diagram_chicchic}
\end{figure}

It was shown in \cite{LS2015,LLP2016} that the $\chi_c J/\psi$ pair production
is possible only at $O(\alpha_s^5)$, while forbidden at $O(\alpha_s^4)$
due to $C$ parity conservation.
In contrast, the production of $\chi_c(J_1) \chi_c(J_2)$, see Fig.\ref{fig:diagram_chicchic},
is possible already at the $O(\alpha_s^4)$ order. 

Of special importance for us is the fact that $\chi_c \chi_c$ states are
produced by the crossed-channel one-gluon exchange mechanism. 
This implies that the production amplitudes are flat as a function
of $g^*g^*$ center of mass energy, which implies broad distributions
in the rapidity distance $\Delta y$ between the produced 
$\chi_c$-mesons.

According to our knowledge this contribution was not discussed so far
in the literature. There was, however,
some calculations for $\chi_c \chi_b$ production \cite{protvino}.

We consider the gluon-gluon fusion mechanism shown diagramatically 
in Fig.\ref{fig:diagram_chicchic}.
There are altogether six possible combinations of pair production
of $\chi_c(0), \chi_c(1), \chi_c(2)$ quarkonia. 

In order to calculate the subprocess amplitudes,
we first turn to the $g^* g^* \to \chi_c(J)$ vertices.

\begin{figure}[!h]
\begin{minipage}{0.32\textwidth}
\centerline{\includegraphics[width=1.0\textwidth]{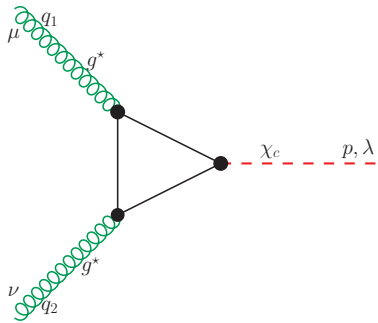}}
\end{minipage}
   \caption{A diagrammatic representation of the 
$g^* g^* \to \chi_c(\lambda)$ vertex being a building block of
corresponding $g^* g^* \to \chi_c(J_1) \chi_c(J_2)$ and discussed
in this section amplitudes.
 }
\label{fig:diag_gg_chic}
\end{figure}

\subsubsection{The $g^* g^* \to \chi_c(J)$ vertices}

The $g^* g^* \to \chi_c$ vertices with off-shell gluons 
(see Fig.\ref{fig:diag_gg_chic}) are building blocks of the 
elementary $g^* g^* \to \chi_c(J_1) \chi_c(J_2)$ amplitudes.

Here we follow the general rules of NRQCD as explained e.g. in 
\cite{Guberina:1980dc,Cho:1995ce,Cho:1995vh}. 
We restrict ourselves to the color singlet contribution and can write the
amplitude for the production of the $\chi_c(J)$ meson via
the fusion of two gluons as:
 \begin{eqnarray}
 V^{ab}_{\mu \nu}(J,J_z;q_1,q_2) = 4 \pi \alpha_S {{\rm Tr}[t^a t^b] \over \sqrt{N_c}} \, 
 \sqrt{{2\over M}} \sum_{S_z,L_z} \int {d^4 k \over (2 \pi)^3} \delta(k^0 - {\vec{k}^2 \over M}) 
 \Psi_{1,L_z}(\vec{k}) 
 \nonumber \\
 \times \bra{1,S_z; 1,L_z}|\ket{J,J_z} \cdot {\rm{Tr}}[A_{\mu \nu} \Pi_{1,S_z}], 
\label{eq:vertex}
 \end{eqnarray}
following closely the notation of 
\cite{Hagler:2000dd,Pasechnik:2007hm,Pasechnik:2009bq,Pasechnik:2009qc},
where these vertices had been calculated for external reggeized gluons.
Below we will need the amplitudes (\ref{eq:vertex}) for arbitrary off-shell momenta
of gluons, not only the multiregge kinematics as in 
\cite{Hagler:2000dd,Pasechnik:2007hm,Pasechnik:2009bq,Pasechnik:2009qc}.
There is however no additional difficulty related with this. As we concentrate on
the color-singlet mechanism, the three-gluon coupling does not enter and we really deal with a QED
problem.
Consequently the amplitudes (\ref{eq:vertex}) fulfill the QED-like gauge invariance conditions:
\begin{eqnarray}
  q_1^\mu V^{ab}_{\mu \nu}(J,J_z;q_1,q_2) = 0, q_2^\nu  V^{ab}_{\mu \nu}(J,J_z;q_1,q_2) = 0.
\end{eqnarray}
The calculation proceeds as follows.  

The $g^* g^* \to Q \bar Q$ amplitude is (up to factors)
\begin{eqnarray}
 A_{\mu \nu} = \gamma_\mu {\hat p_Q - \hat q_1 + m_Q \over (p_Q - q_1)^2 - m_Q^2} \gamma_\nu + 
 \gamma_\nu {\hat p_Q - \hat q_2 + m_Q \over (p_Q - q_2)^2 - m_Q^2} \gamma_\mu \, . 
\end{eqnarray}
We parametrize
\begin{eqnarray}
 p_Q = {P \over 2 } + k \, , \, p_{\bar Q} = {P \over 2} - k \, ,
\end{eqnarray}
In spectroscopic notation, the $\chi_c$ mesons are $^{2S+1}L_J = ^3 P_J$ states,
where $J=0,1,2$. Therefore the spinorial part of the wavefunction is
an $S=1$ spin triplet state, and the relevant projector can be
written as
\begin{eqnarray}
 \Pi_{S=1,S_z} &=& {1 \over 2 \sqrt{2}  m_Q} \, \Big({\hat P \over 2} - \hat k - m_Q \Big) \hat \epsilon(S_z)  
 \Big({\hat P \over 2} + \hat k + m_Q \Big). 
\end{eqnarray}
Now, for the $P$-wave states, we should expand the product $A_{\mu \nu}\hat \Pi_{S=1,S_z}$ in
(\ref{eq:vertex}) to the first order in $k$. In fact the Taylor expansion for $P$-waves starts
from the term linear in $k$:
\begin{eqnarray}
  {\rm Tr}[A_{\mu \nu} \Pi_{1,S_z}] \rightarrow  k_\alpha \cdot {\partial \over \partial k_\alpha}  
  {\rm Tr}[A_{\mu \nu} \Pi_{1,S_z}]\Big|_{k=0} \, .
\end{eqnarray}
Then, the integration over relative momentum $k$ reduces to the integral
\begin{eqnarray}
\int {d^3 \vec{k} \over (2 \pi)^3} k^\alpha \Psi_{1,L_z}(\vec{k}) = 
-i \sqrt{3 \over 4 \pi} \, R'(0) \cdot \epsilon^\alpha(L_z) \, .
\end{eqnarray}
Here $R'(0)$ is the derivative of the radial wavefunction at the (spatial) origin.
 
For convenience, we introduce
\begin{eqnarray} 
T_{\mu \nu} (q_1,q_2;J,J_z) \equiv 
{\sqrt{2} M \over 8} \, \sum_{S_z,L_z} \bra{1,S_z; 1,L_z}|\ket{J,J_z} \epsilon_\alpha(L_z) 
\cdot {\partial \over \partial k_\alpha}  {\rm Tr}[A_{\mu \nu} \Pi_{1,S_z}]\Big|_{k=0},
\end{eqnarray}
so that our  gluon-gluon fusion vertices take the form
\begin{eqnarray}
 V^{ab}_{\mu \nu}(J,J_z;q_1,q_2) = -i \, 4 \pi \alpha_S \,   
 \delta^{ab} \, {2R'(0) \over \sqrt{\pi N_c M^3}} \, \sqrt{3} 
 \cdot T_{\mu \nu}(J,J_z;q_1,q_2) \, ,
\end{eqnarray}
Performing the relevant Dirac-traces, we obtain 
the explicit expressions for $T_{\mu \nu}(J,J_z;q_1,q_2)$:
\begin{enumerate}
 \item scalar, $J=0$:
\begin{eqnarray}
 T_{\mu \nu}(0,0;q_1,q_2) &=& {1 \over \sqrt{3}} \,  {M^2 \over (2 q_1 \cdot q_2)^2} 
 \, \Big\{ g_{\mu \nu} \Big( 6(q_1 \cdot q_2) - q_1^2 -q_2 ^2 +
 {(q_2^2 - q_1^2)^2 \over M^2}\Big) \nonumber \\
 &+& q_{1 \mu} q_{2\nu} \, 2\Big( {q_1^2 + q_2^2 \over M^2} -1 \Big) \, 
 + q_{2 \mu} q_{1 \nu} \, 2\Big( {q_1^2 + q_2^2 \over M^2} -3 \Big)
 \nonumber\\
 &+& q_{1 \mu} q_{1\nu} {4 q_2^2 \over M^2} + q_{2 \mu} q_{2 \nu} {4 q_1^2 \over M^2} \Big\}
 \label{eq:scalar}
\end{eqnarray}
 \item axial vector, $J=1$:
\begin{eqnarray}
 T_{\mu \nu} (1,J_z;q_1,q_2) &=& {i \over \sqrt{2}M} {1 \over (q_1 \cdot q_2)} \Big\{ 
  (q_1^2 - q_2^2) 
 \epsilon_{\mu \nu \alpha \beta} (q_1+ q_2)^\alpha \epsilon^{\beta}(J_z) 
 \nonumber \\
 &+& {q_1^2 + q_2^2 \over  (q_1 \cdot q_2)}   (a_\mu q_{1\nu} - a_\nu q_{2\mu})
 + 2 ( a_\nu q_{1\mu} - a_\mu q_{2\nu}) \Big\} \, 
 \label{eq:axial}
\end{eqnarray}
with
\begin{eqnarray}
 a_\mu =  \epsilon_{\mu \rho \alpha \beta} q_1^\rho q_2^\alpha \epsilon^{\beta}(J_z) \, .
\end{eqnarray}
\item tensor, $J=2$:
\begin{eqnarray}
 T_{\mu \nu}(2,J_z;q_1,q_2) &=& {- M^2 \over (2q_1 \cdot q_2)^2} 
 \Big\{ - g_{\mu \nu} (q_2 -q_1)^\alpha (q_2 - q_1)^\beta 
 \epsilon_{\alpha \beta}(J_z)  + 4 (q_1 \cdot q_2) \epsilon_{\mu \nu}(J_z) 
 \nonumber \\
 &+& 2 (q_2 -q_1)^\alpha \epsilon_{\alpha \nu}(J_z) q_{2\mu} 
 - 2 (q_2 - q_1)^\alpha \epsilon_{\alpha \mu}(J_z) q_{1\nu} \Big\} \, ,
 \label{eq:tensor}
\end{eqnarray}
where 
$\epsilon_{\mu \nu}(J_z) = \sum_{m_1,m_2} \bra{2,J_z}|\ket{1,m_1,1,m_2} \epsilon_\mu(m_1) \epsilon_\nu(m_2)$ 
is the polarization tensor of the $J=2$ state.
\end{enumerate}
Notice, that 
\begin{eqnarray}
 2(q_1 \cdot q_2) = M^2 - q_1^2 - q_2^2 \, ,
\end{eqnarray}
and as gluons are always spacelike $q_i^2 <0$, the denominators of eqs (\ref{eq:scalar}
,\ref{eq:axial}, \ref{eq:tensor}) are always finite.

Besides the QED-like gauge invariance condition, these amplitudes also fulfill the Bose-symmetry
\footnote{Notice that it does not mean that $T_{\mu \nu}$ is a symmetric tensor, as the 
results presented in \cite{protvino} (which violate the gauge invariance condition).}
\begin{eqnarray}
  T_{\mu \nu}(J,J_z;q_1,q_2) =   T_{\nu \mu}(J,J_z;q_2,q_1) \, .
\end{eqnarray}
A comment on the $J=1$ axial vector is in order.
Here the Landau-Yang theorem forbids the decay of the $\chi_c(1)$ 
into $\gamma \gamma$ or $gg$, and likewise its production
through fusion of on-shell photons or gluons.
Indeed, in the limit $q_1^2 \to 0, q_2^2 \to 0$,
we have 
\begin{eqnarray}
 T_{\mu \nu} (1,J_z;q_1,q_2) \propto a_\nu q_{1\mu} - a_\mu q_{2\nu}  \, ,
\end{eqnarray}
which vanishes, when contracted with the polarization vectors
of on-shell photons/gluons
\begin{eqnarray}
 \epsilon_1^\mu \epsilon_2^\nu ( a_\nu q_{1\mu} - a_\mu q_{2\nu} ) = 0,
\end{eqnarray}
as required by the Landau-Yang theorem.
\subsubsection{The $g^* g^* \to \chi_c(J_1) \chi_c(J_2)$ amplitudes}
Now we wish to discuss the elementary 
$g^* g^* \to \chi_c(J_1) \chi_c(J_2)$ amplitudes, which can be obtained from the
building blocks discussed above.

\begin{figure}[!h]
\includegraphics[width=4cm]{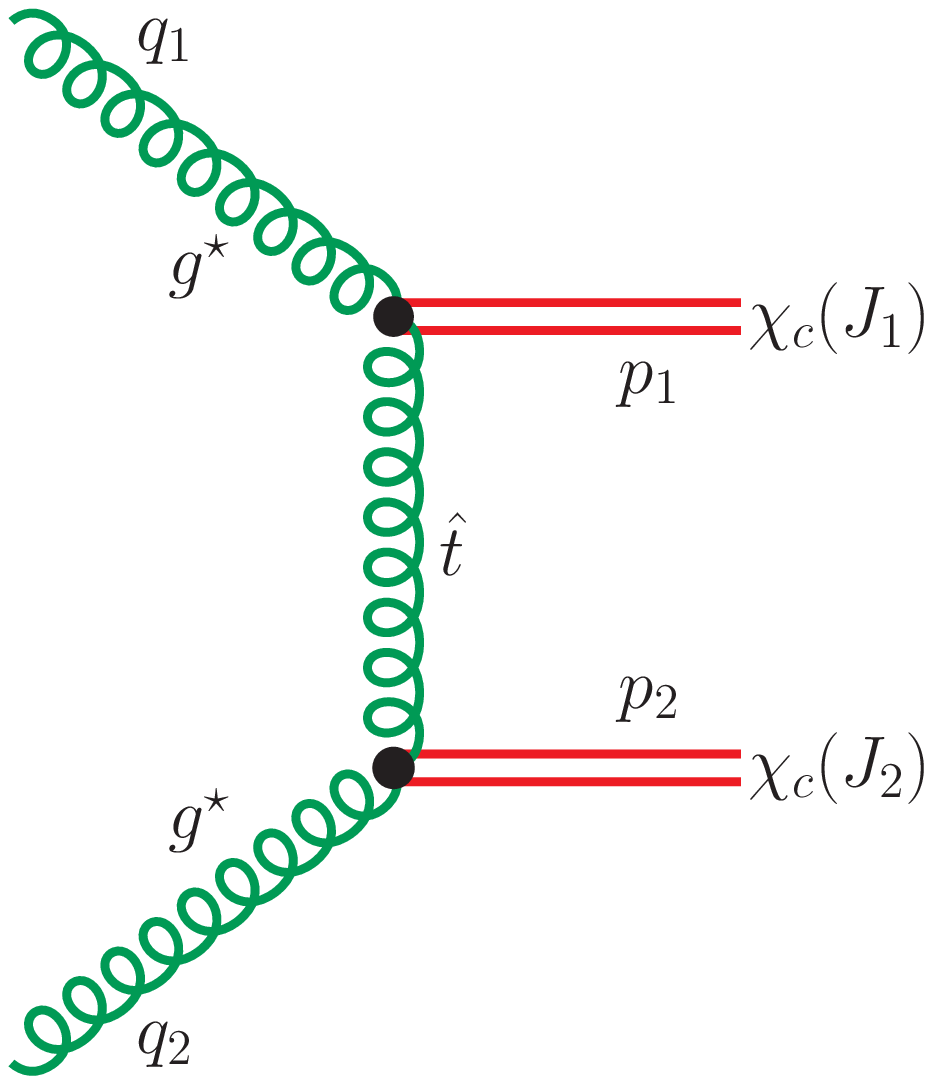}
\includegraphics[width=4cm]{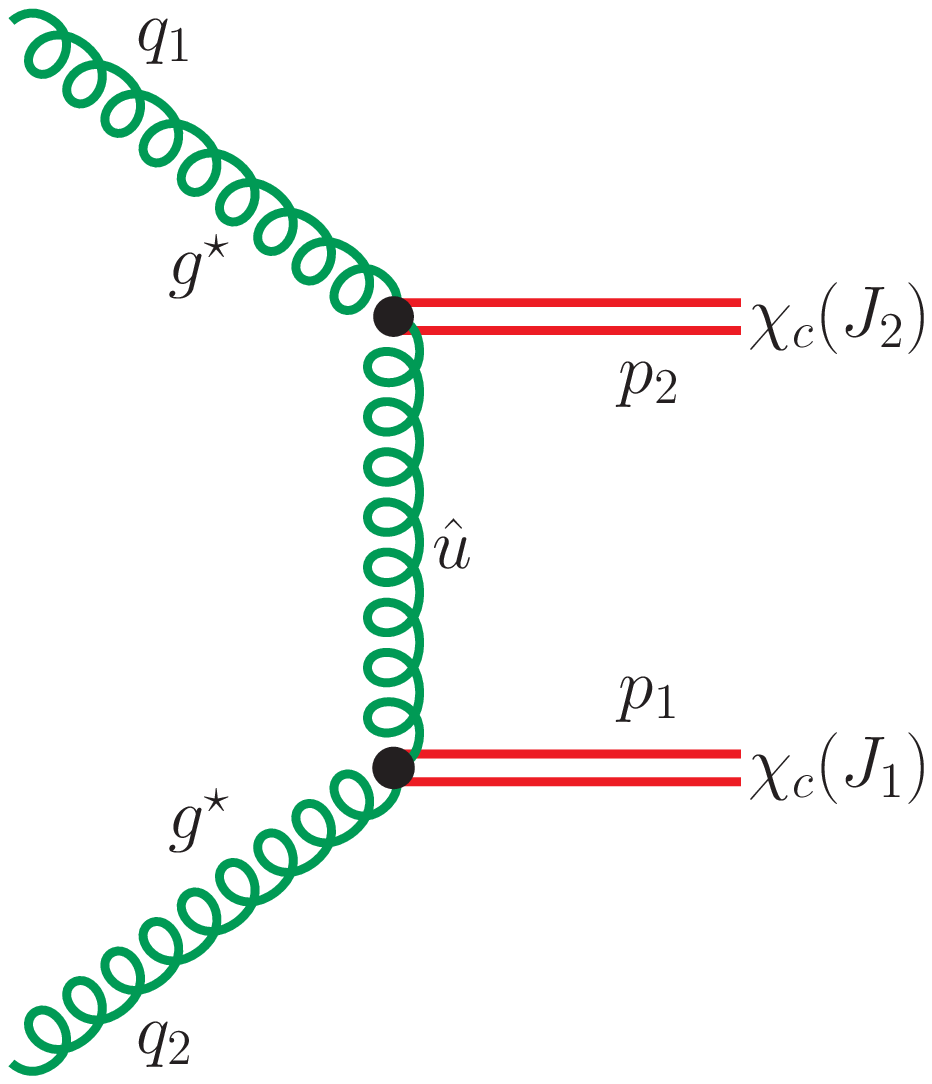}
   \caption{A diagrammatic representation of the 
generic $g^* g^* \to \chi_c(J_1) \chi_c(J_2)$ $t$-channel (left) 
and $u$-channel (right) amplitudes.
 }
\label{fig:chicchic_amplitudes}
\end{figure}

In all cases there are two diagrams ($t$ (left) and $u$ (right)
in Fig.\ref{fig:chicchic_amplitudes}).

We can write the Feynman amplitudes corresponding to these diagrams as
\begin{eqnarray}
 {\cal M}^{ab}_{\mu \nu}(J_1,J_{1z},J_2,J_{2z}) &=&  V^{ac}_{\mu \alpha}(J_1,J_{1z};q_1,p_1-q_1)
 {-g^{\alpha \beta} \delta_{cd} \over \hat t} V^{db}_{\beta \nu}(J_2,J_{2z};p_2-q_2,q_2) \nonumber\\
 &+& V^{ac}_{\mu \alpha}(J_2,J_{2z};q_1,p_2-q_1)
 {-g^{\alpha \beta} \delta_{cd} \over \hat u} V^{db}_{\beta \nu}(J_1,J_{1z};p_1-q_2,q_1) \, ,
\label{eq:chi_chi}
 \end{eqnarray}
where $\hat t = (p_1 - q_1)^2 = (p_2-q_2)^2 \, , \, \hat u = (p_2 - q_1)^2 = (p_1 - q_2)^2$.
These amplitudes are infrared finite and gauge invariant.

To obtain the $k_T$-factorization amplitude one should contract (\ref{eq:chi_chi}) with the
polarization vectors of off-shell gluons
\begin{eqnarray}
 e_{1 \mu} = {q_{1T \mu} \over |\vec{q}_{1T}| } \, , \, 
 e_{2 \nu} = {q_{2T \nu} \over |\vec{q}_{2T}| } \, .
\end{eqnarray}
Because of the QED-like Ward identities of the gluon fusion vertices, 
these polarization vectors are equivalent to the more common Gribov's polarizations 
$n^+_\mu, n^-_\nu$, for incoming gluons
in the high-energy kinematics $q_{1\mu} = q_1^+ n^+_\mu + q_{1T \mu}$, 
 $q_{2\nu} = q_1^+ n^-_\nu + q_{2T \nu}$.

In the nonrelativistic QCD approach the cross section for  
$\chi_c$ pair production is proportional to $|R'(0)|^4$.
The result is therefore extremely sensitive to the precise value
of the wave function derivative at the origin.
In our opinion the best estimate of the parameter can be obtained from:
\begin{equation}
\Gamma(\chi_c(0^+) \to \gamma \gamma) = 
{27 e_c^4 \alpha_{\rm em} \over m_c^4} |R'(0)|^2 \; .
\end{equation}
From the experimental value of the diphoton decay width \cite{PDG}
one obtains for the $\chi_c$ P-wave function squared 
\begin{equation}
|R'(0)|^2 = 0.042 \; {\rm GeV}^5.
\end{equation}

In the following the $\chi_c(J_1) \chi_c(J_2)$ cross section is
calculated within the $k_T$-factorization approach including
off-shell matrix elements for the $g^* g^* \to \chi_c(J_1) \chi_c(J_2)$
subprocess and modern unintegrated gluon distributions.

It is well known that about 30 \% of prompt single $J/\psi$ production
originates from radiative decays $\chi_c \to J/\psi + \gamma$
with branching fractions:
${\rm{Br}}(\chi_c(0) \to J/\psi \gamma) =  1.26 \pm 0.06 \% $,
${\rm{Br}}(\chi_c(1) \to J/\psi \gamma) = 33.9 \pm 1.2 \%$,
${\rm{Br}}(\chi_c(2) \to J/\psi \gamma) = 19.2 \pm 0.7 \%$ \cite{PDG}.
Obviously, regarding feed down into the $J/\psi J/\psi$ channel only
the $\chi_c(1) \chi_c(1)$,
$\chi_c(1) \chi_c(2)$ and $\chi_c(2) \chi_c(2)$ states could give potentially
important contributions. The details depend, however, on corresponding matrix elements
and cross sections for the $\chi_c(J_1) \chi_c(J_2)$ production.


The cross section for $p p \to \chi_c(J_1) \chi_c(J_2)$
is calculated in the $k_T$-factorization approach.
The corresponding differential cross section for the production
of $\chi_c(i) \chi_c(j)$ states, where $i$ and $j$ run through $0,1,2$
can be written as:
\begin{eqnarray}
&&\frac{d \sigma(p p \to \chi_c(i) \chi_c(j) X)}
{d y_1 d y_2 d^2 {\vec p}_{1T} d^2 \vec{p}_{2T}}
 = 
\frac{1}{16 \pi^2 (x_1 x_2 s)^2} {1 \over 1+ \delta_{ij}} \int \frac{d^2 \vec{q}_{1T}}{\pi \vec{q}_{1T}^2} 
\frac{d^2 \vec{q}_{2T}}{\pi \vec{q}_{2T}^2} 
\overline{|{\cal M}_{g^{*} g^{*} \rightarrow \chi_c(i) \chi_c(j)}^{\rm{off-shell}}|^2} 
\nonumber \\
&& \times \;\; 
\delta^{(2)} \left( \vec{q}_{1T} + \vec{q}_{2T} - \vec{p}_{1T} - \vec{p}_{2T} \right)
{\cal F}(x_1,\vec{q}_{1T}^2,\mu_{F}^{2}) 
{\cal F}(x_2,\vec{q}_{2T}^2,\mu_{F}^{2}) \; .
\label{kt_fact_gg_jpsijpsi}
\end{eqnarray}
The unintegrated gluon distribution ${\cal F}(x_1,\vec{q}_{1T}^2,\mu_{F}^{2})$ is related to the
collinear one through
\begin{eqnarray}
x g(x,\mu_F^2) = \int^{\mu^2_F} {d \vec{q}^2_T \over\vec{q}^2_T  } \, {\cal F}(x,\vec{q}_{1T}^2,\mu_{F}^{2})\, ,
\end{eqnarray}
and the off-shell matrix element is obtained as
\begin{eqnarray}
 \overline{|{\cal M}_{g^{*} g^{*} \rightarrow \chi_c \chi_c}^{\rm{off-shell}}|^2}
 ={1 \over (N_c^2-1)^2} \sum_{a,b,J_{1z},J_{2z}} 
 |e_{1\mu} e_{2\nu} {\cal{M}}^{ab}_{\mu \nu}(J_1,J_{1z},J_2,J_{2z})|^2
 \; .
\end{eqnarray}
The longitudinal momentum fractions $x_1$ and $x_2$ are calculated from $\chi_c$'s 
transverse masses $m_{Ti} = \sqrt{m_c^2 + \vec{p}_{iT}^2}$ and rapidities:
\begin{eqnarray}
x_1 &=& {m_{T1} \over \sqrt{s} } e^{y_1} +  {m_{T2} \over \sqrt{s} } e^{y_2} \, , \nonumber \\
x_2 &=& {m_{T1} \over \sqrt{s} } e^{-y_1} +  {m_{T2} \over \sqrt{s} } e^{-y_2}  \, .
\end{eqnarray}
\subsection{Results for $\chi_c(J_1) \chi_c(J_2)$ production}

We start presentation of our results by showing integrated cross
sections.
As an example in Table \ref{table:cross_sections} 
we show cross section in a broad range of $\chi_c$ rapidities.
We used an unintegrated gluon distribution constructed from the KMR 
prescription \cite{Kimber:2001sc} based on the MSTW2008 collinear NLO
gluon distribution \cite{MSTW08}.
For the renormalization scales $\mu^2_{r1},\mu^2_{r2}$  of the running coupling
and factorization scales $\mu^2_{F1},\mu^2_{F2}$ entering the unintegrated
gluon distribution, we choose
\begin{eqnarray}
\mu_{r1}^2 &=& \mu^2_{F1} = {\rm max}\{m_c^2, |\vec{q_{1T}}|^2\} \, ,\nonumber \\
\mu_{r2}^2 &=& \mu^2_{F2} = {\rm max}\{m_c^2, |\vec{q_{2T}}|^2\} \ ,
\end{eqnarray}
where these scales refer to the running coupling/gluon distribution coupling
to gluon $q_1$ or $q_2$ respectively.
We refrain from a detailed analysis of dependence on the factorization scale,
the distributions shown below simply serve to get an impression of the
salient features of the production mechanism.
A more detailed analysis, including theoretical errors will be given in a future work
\cite{paperJPsi}, where we will address the feeddown into the $J/\psi J/\psi$ channel.

\begin{table}
\begin{center}
\begin{tabular}{|c|c|c|c|}
\hline
             &  $\chi_c(0)$  &  $\chi_c(1)$  &  $\chi_c(2)$  \\
\hline
$\chi_c(0)$  &  1.32       &   1.71         &   4.24       \\
$\chi_c(1)$  &  ....       &   0.84         &   2.88       \\
$\chi_c(2)$  &  ....       &   ....         &   3.45     \\
\hline
\end{tabular}
\end{center}
\caption{Cross sections in nb for production of different combinations
	of $\chi_c(J_1) \chi_c(J_2)$ dimeson states
	for -8 $< y_1, y_2 <$ 8 at $\sqrt{s}$ = 8 TeV.
	The numbers are obtained in the $k_T$-factorization approach.
	We used an unintegrated gluon distribution constructed from the KMR 
	prescription \cite{Kimber:2001sc} based on the MSTW2008 collinear NLO
	gluon distribution \cite{MSTW08}.
	In all cases the gauge invariant matrix elements discussed in the
	present paper were used.}
\label{table:cross_sections}
\end{table}


There are six independent cross sections related to the different spin 
combinations (see Table \ref{table:cross_sections}).
We see that the cross sections for different spin combinations
are of the same order of magnitude. 

\begin{figure}[!h]
	\includegraphics[width=.4\textwidth]{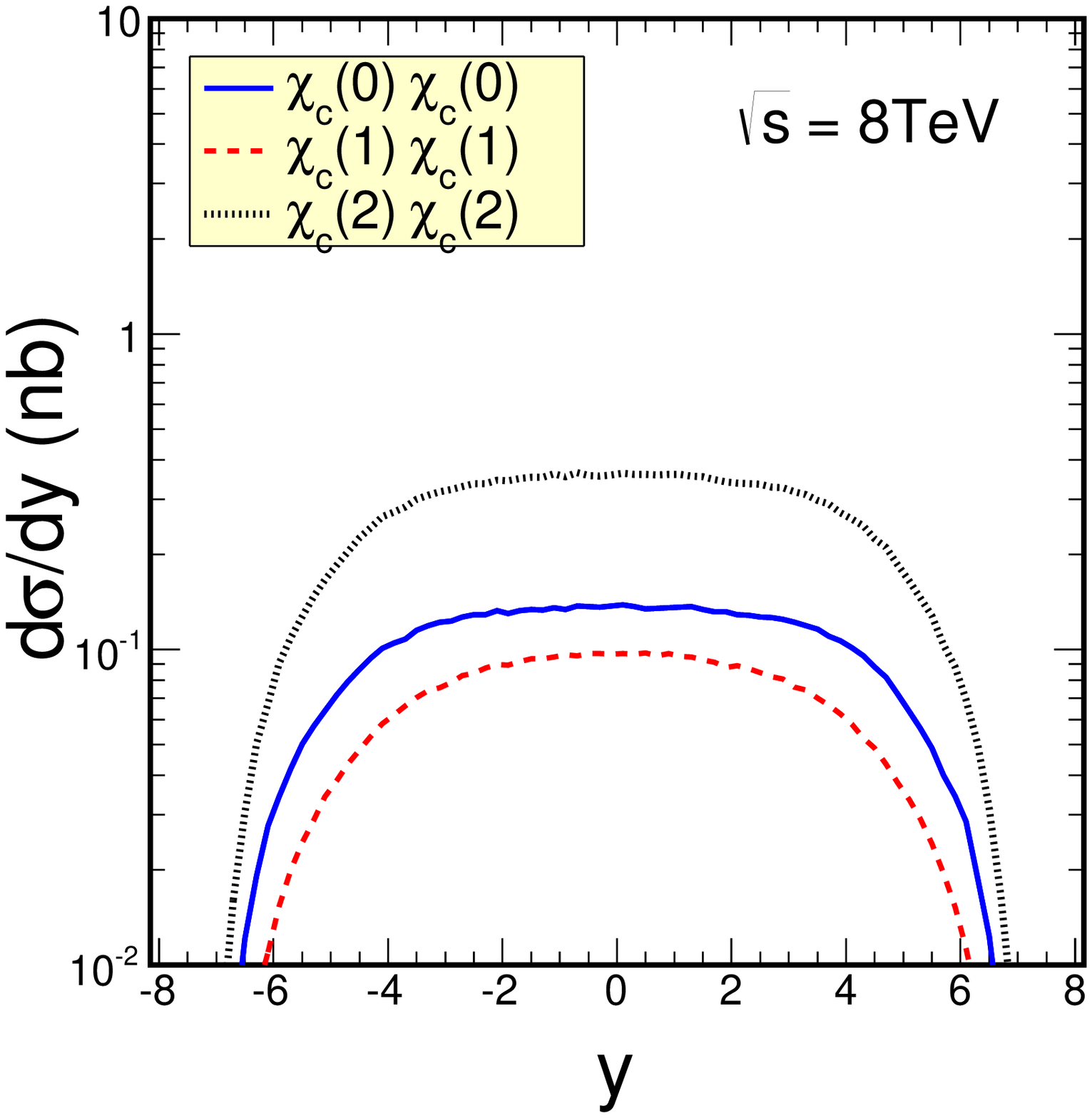}
	\includegraphics[width=.4\textwidth]{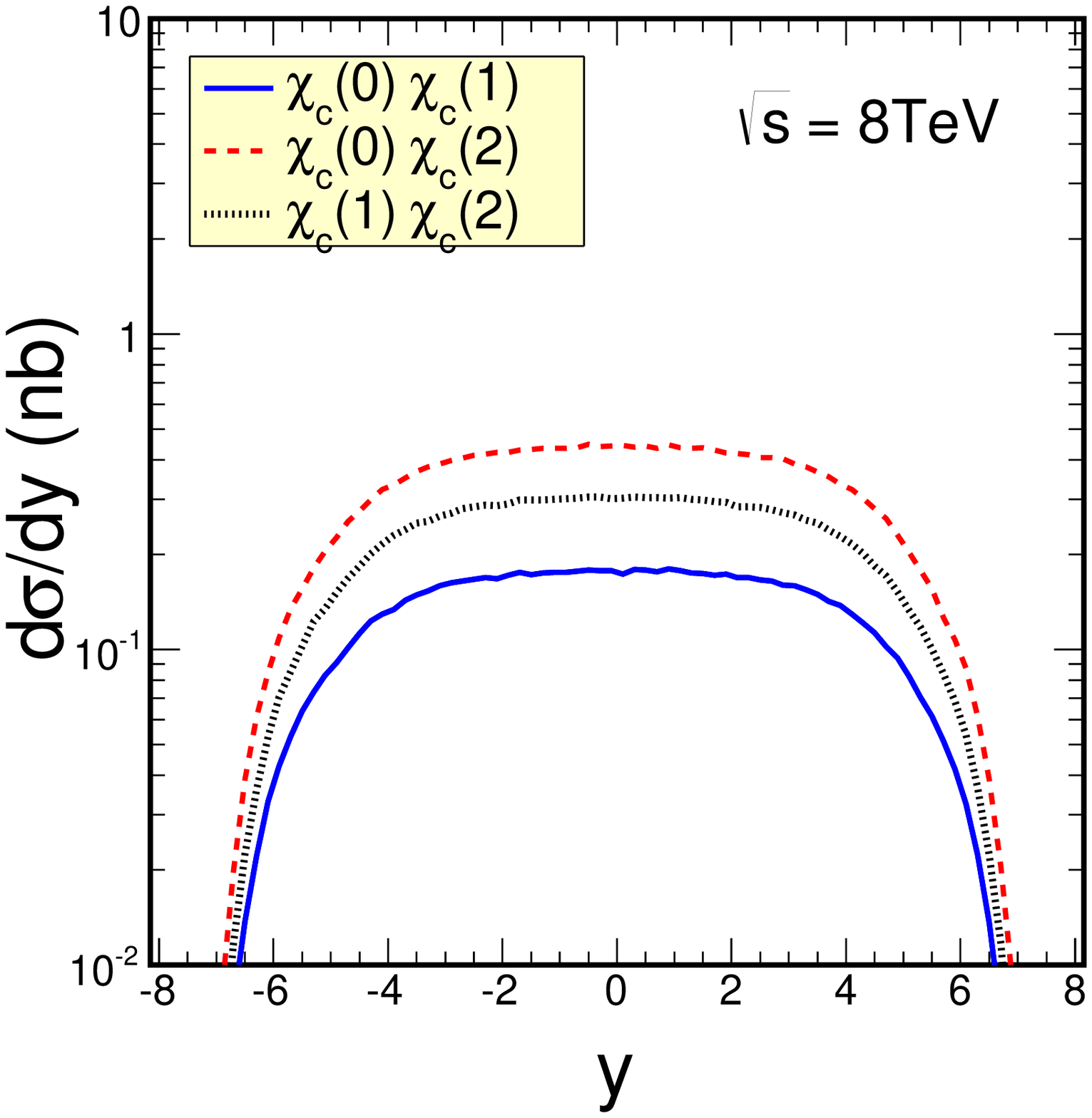}
	\includegraphics[width=.4\textwidth]{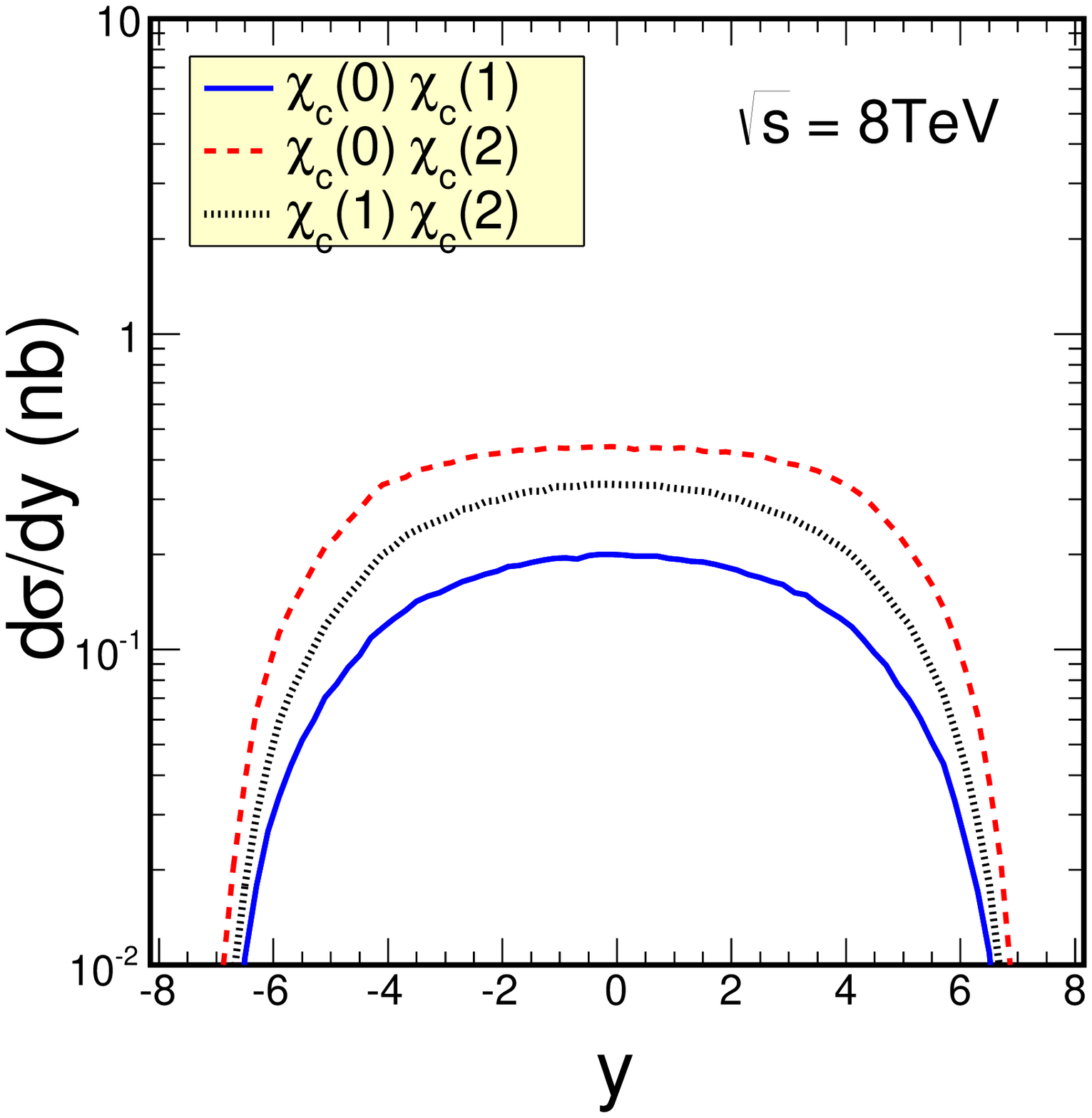}
	\caption{Rapidity distributions of quarkonia for different spin
		combinations.}
	\label{fig:dsig_dy}
\end{figure}

In Fig.\ref{fig:dsig_dy} we show rapidity distributions for
$\chi_c$ mesons for different pair combinations. 
In the left panel we show:
$\chi_c(0) \chi_c(0)$ (solid line), $\chi_c(1) \chi_c(1)$ (dahed line)
and $\chi_c(2) \chi_c(2)$ (dotted line).
In the other panels we show distributions for:
$\chi_c(0) \chi_c(1)$ (solid line), $\chi_c(0) \chi_c(2)$ (dashed line)
and $\chi_c(1) \chi_c(2)$ (dotted line).
In the upper keft panel the distribution of the first listed quarkonium is shown, 
while the distributions of the second listed quarkonium
are shown in the lower panel. Evidently for the nonidentical quarkonia the 
distribution of the first and second meson are not the same.

\begin{figure}[!h]
\includegraphics[width=.4\textwidth]{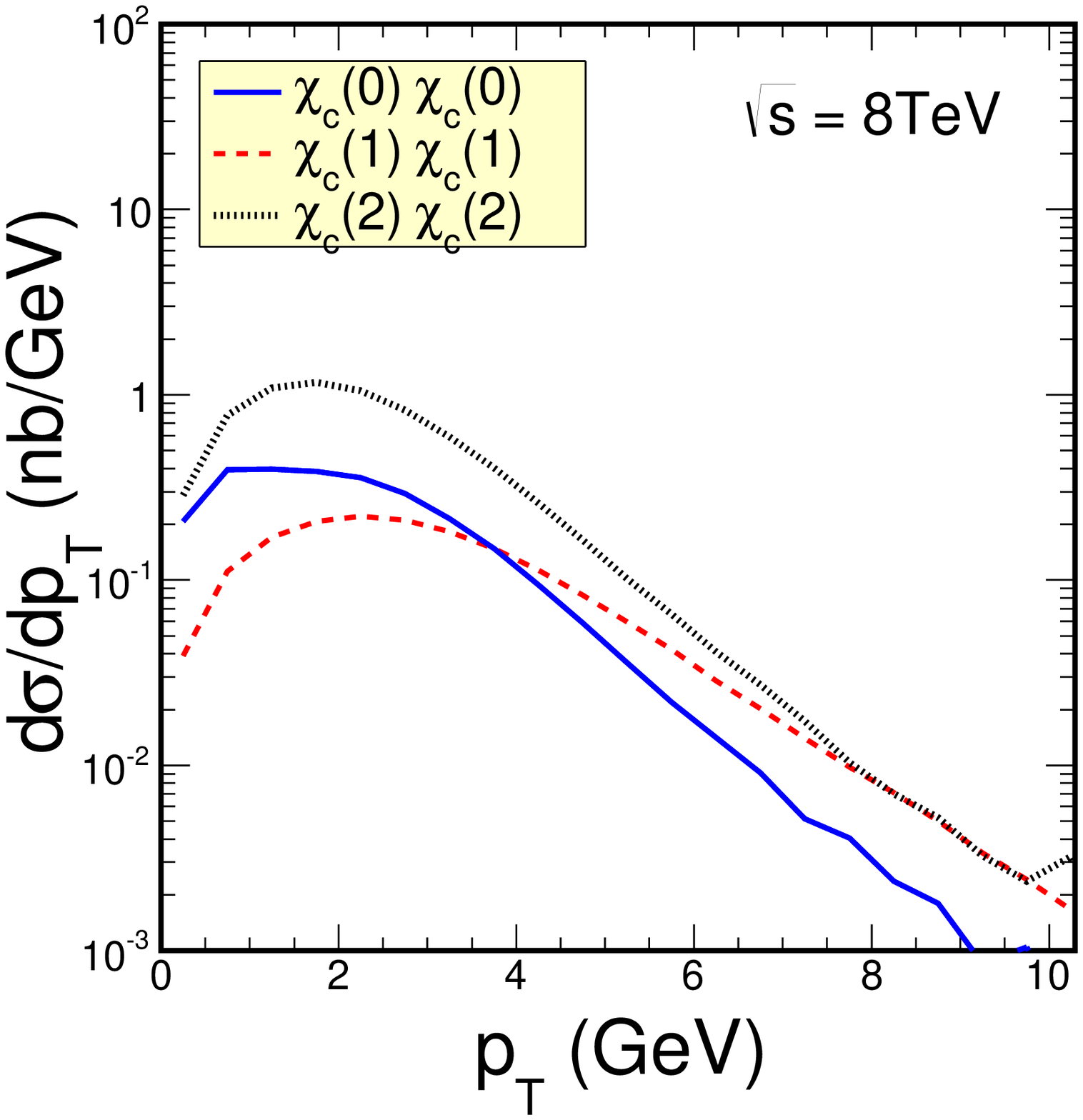}
\includegraphics[width=.4\textwidth]{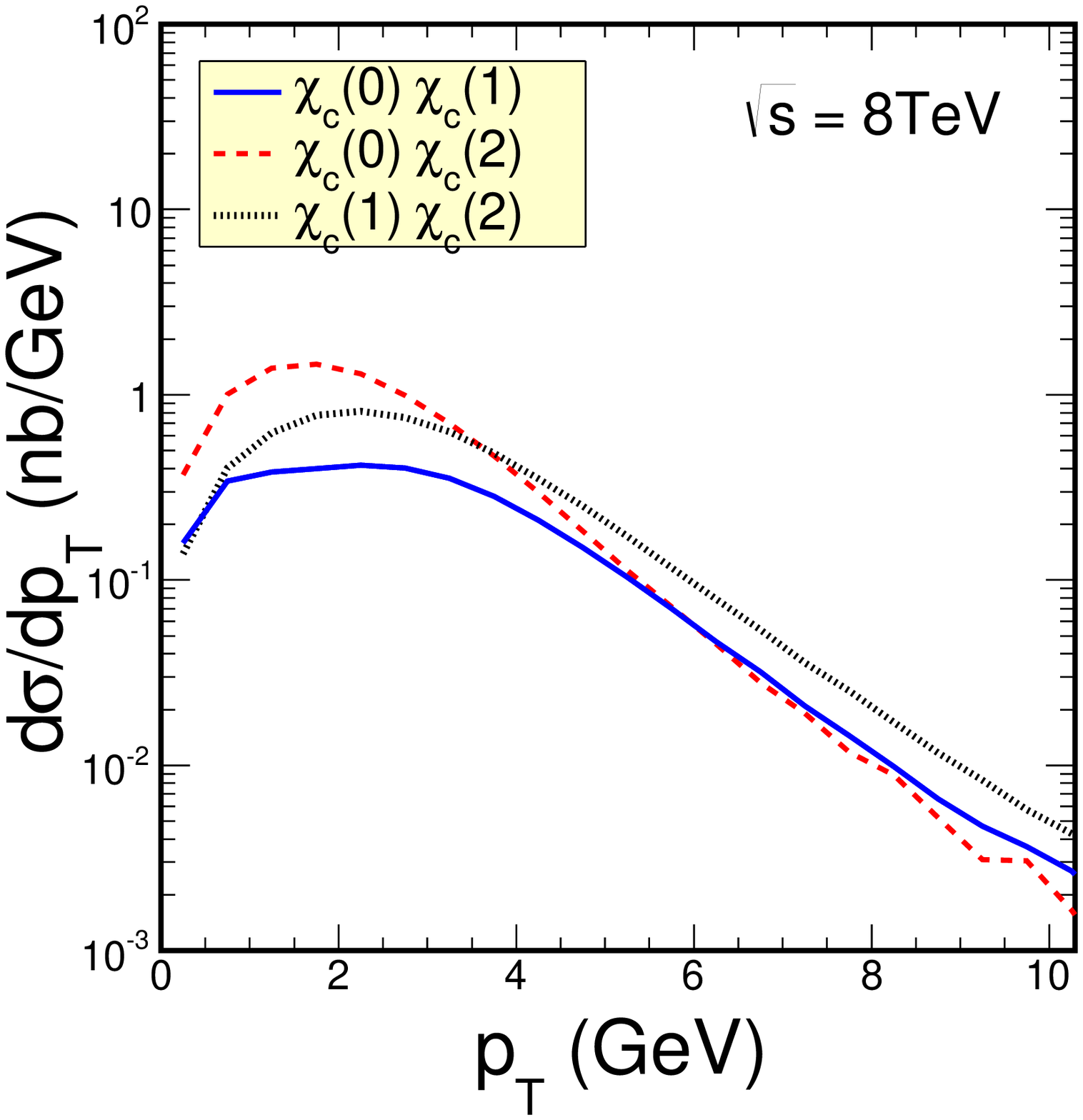}
\includegraphics[width=.4\textwidth]{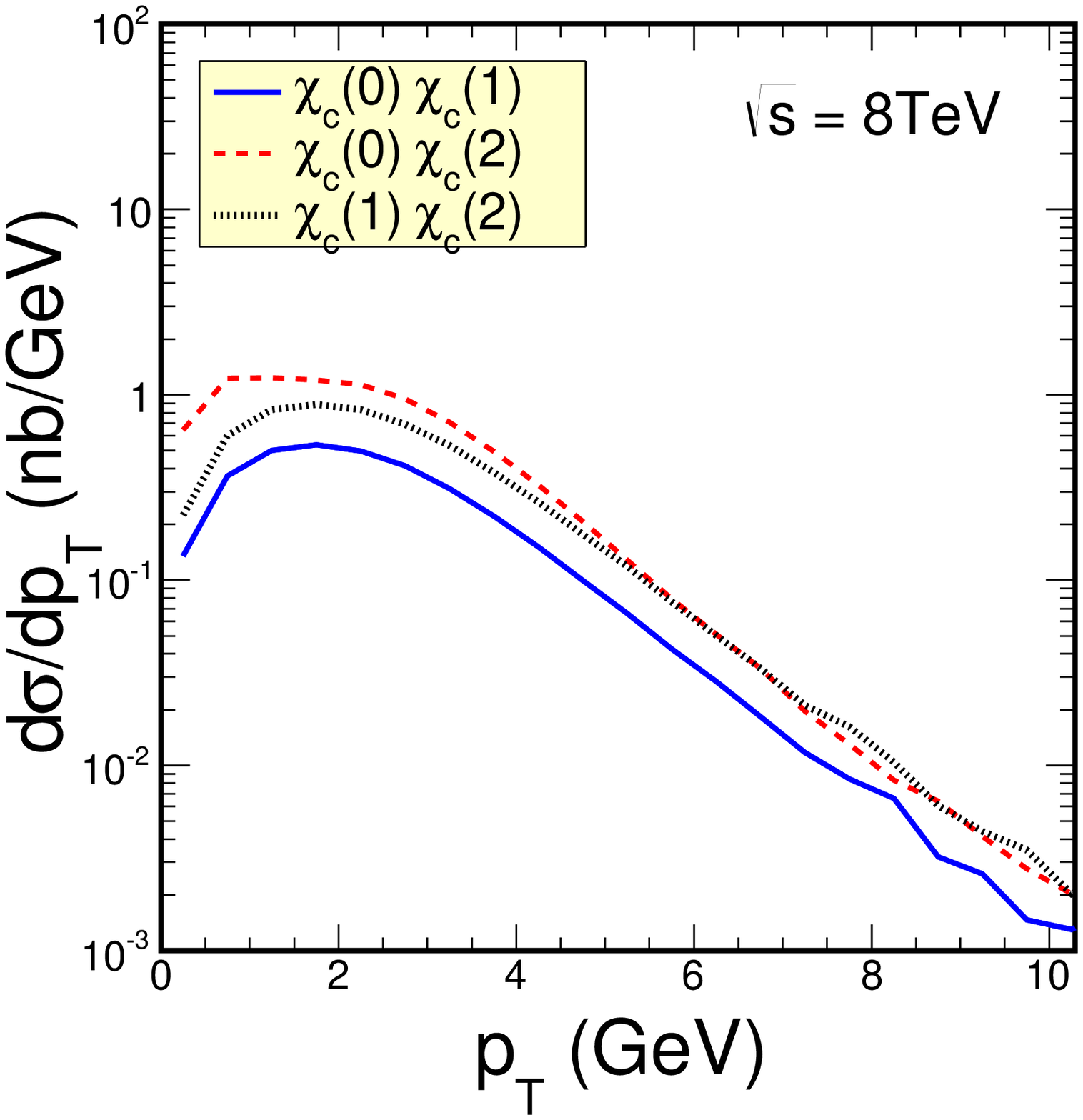}
\caption{Transverse momentum distributions of quarkonia for different 
spin combinations.}
\label{fig:dsig_dpt}
\end{figure}

In Fig.\ref{fig:dsig_dpt} we show similar distributions in quarkonia
transverse momenta. The distributions for $\chi_c(1)$ quarkonia are
less steep than those for the other mesons. This may have important
consequences for large transverse momenta, also for $J/\psi$ pair production
(CDF, ATLAS, CMS), but goes beyond the scope of the present letter.

\begin{figure}[!h]
\includegraphics[width=.4\textwidth]{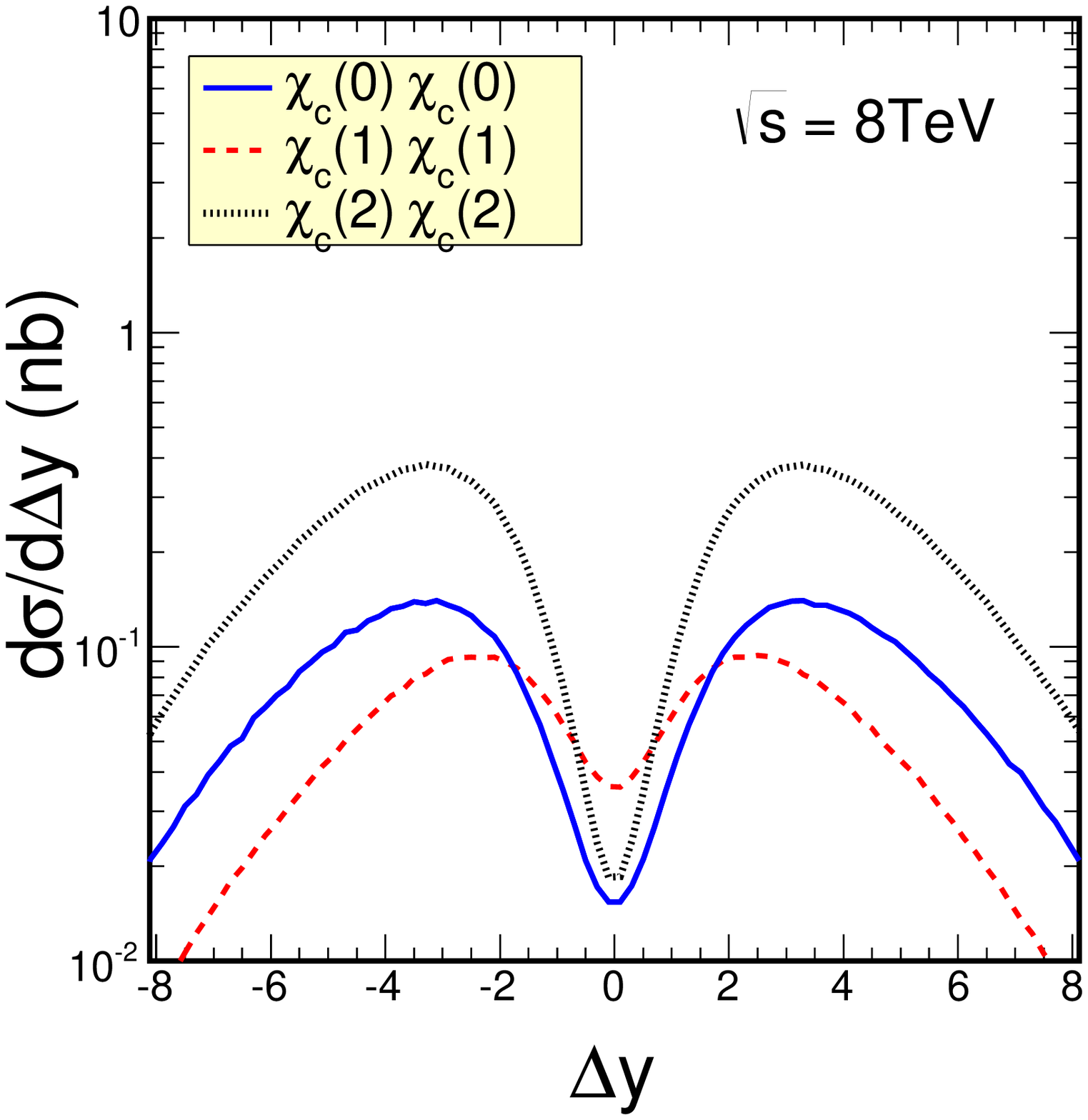}
\includegraphics[width=.4\textwidth]{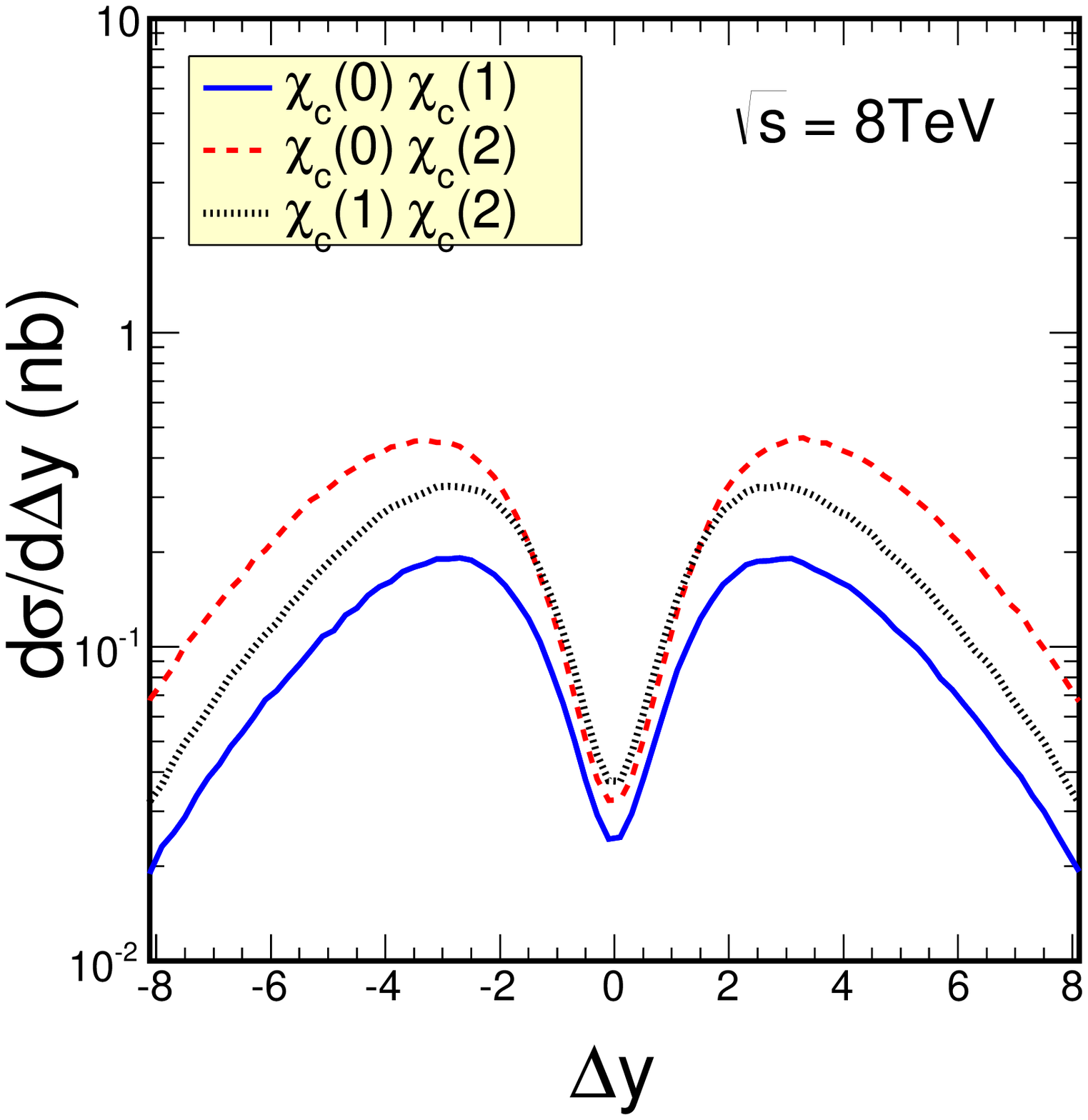}
\caption{Distributions in the rapidity separation between $\chi_c$'s 
for different spin combinations.}
\label{fig:dsig_dydiff}
\end{figure}

The exchange of gluons leads to broad distributions in the difference
of rapidities $\Delta y$ of the two quarkonia, as shown in 
Fig.\ref{fig:dsig_dydiff}. All final states have in common also a rather 
deep dip at $\Delta y =0$. Therefore the $\chi_c$ pair production will be 
potentially important rather for experimental setups that cover a large range in 
rapidities.

\begin{figure}[!h]
\includegraphics[width=.4\textwidth]{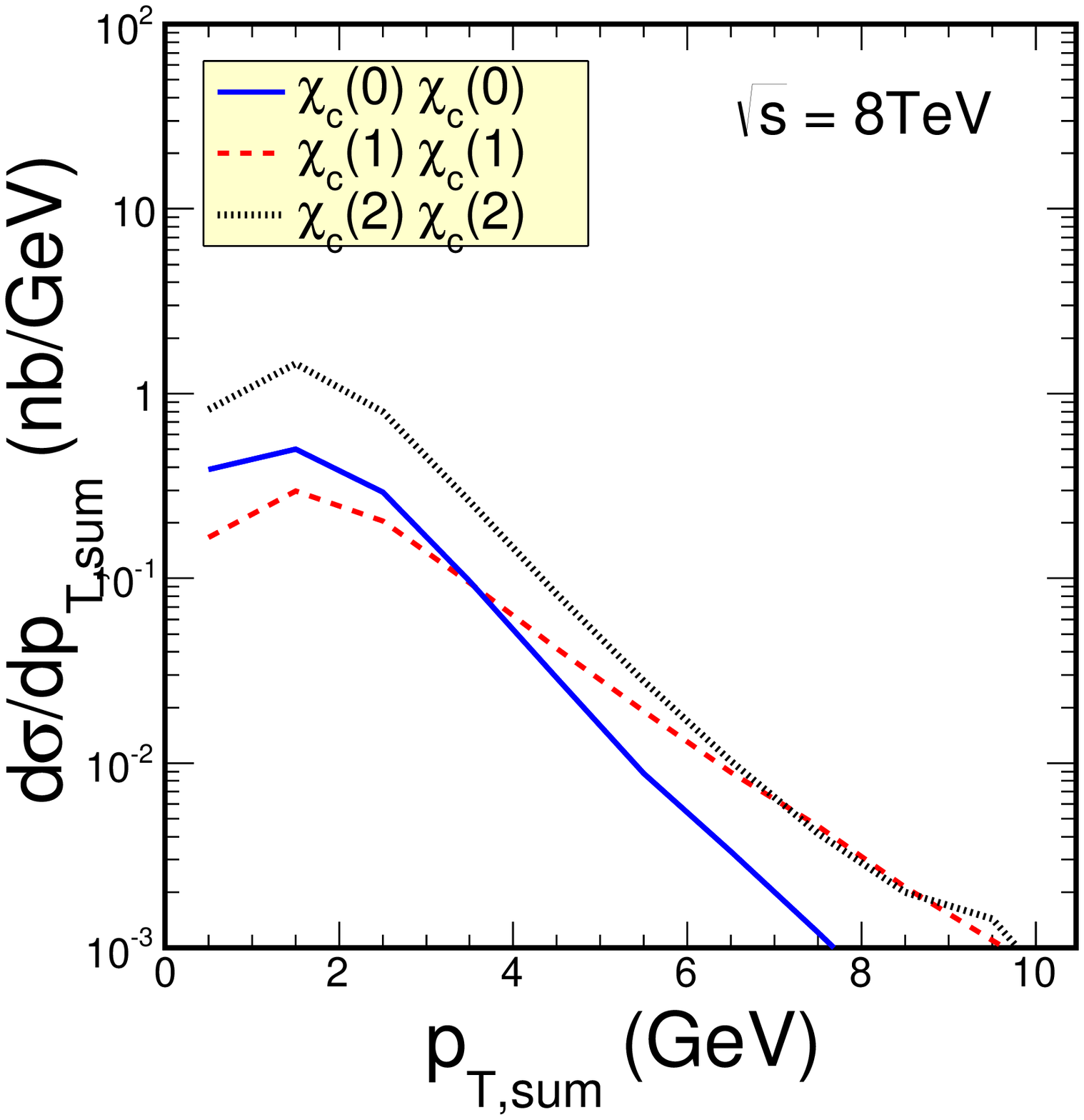}
\includegraphics[width=.4\textwidth]{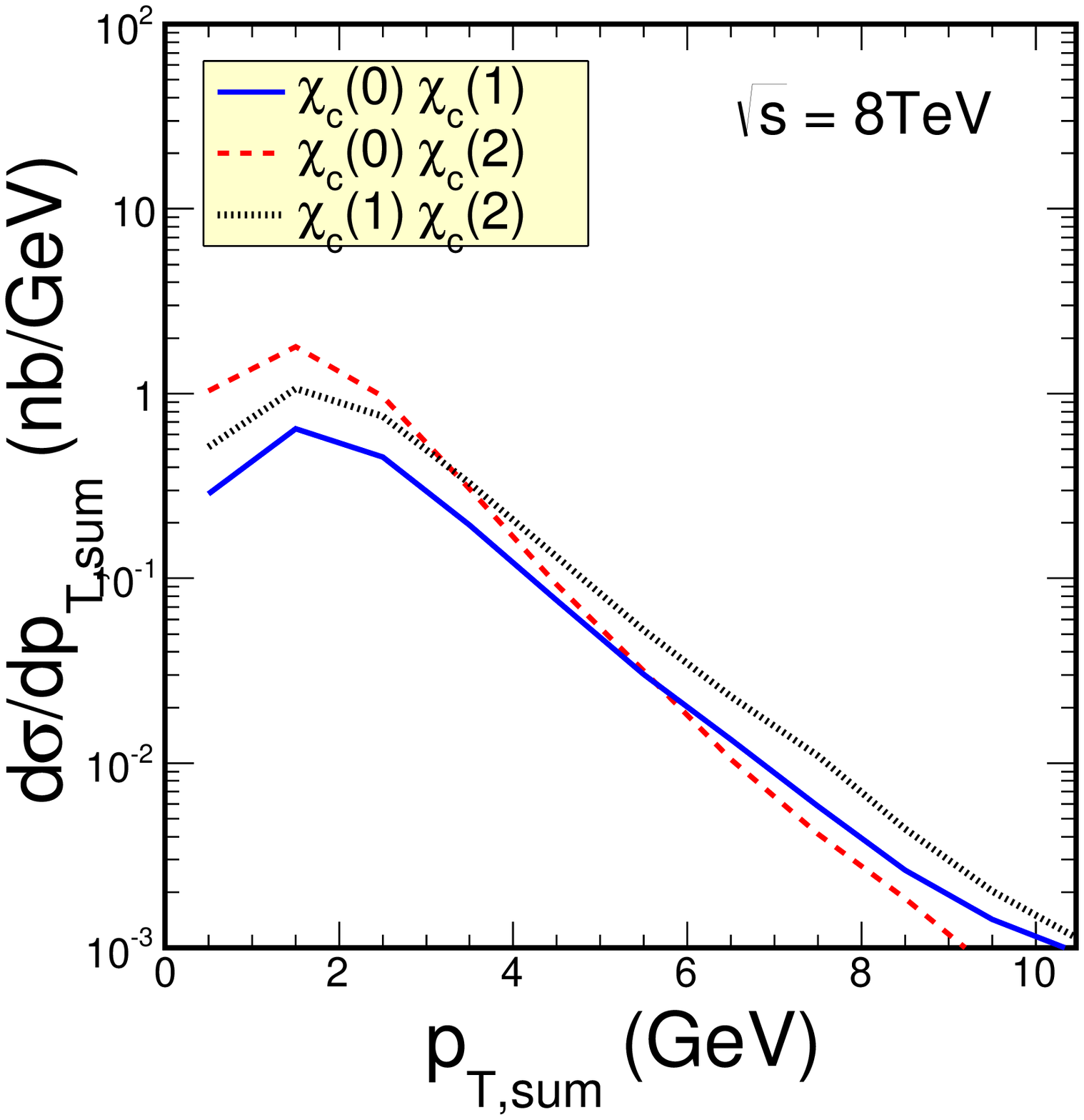}
\caption{Distributions in the transverse momentum of quarkonium pairs 
for different spin combinations.}
\label{fig:dsig_dptsum}
\end{figure}

In calculations based on collinear gluon distributions, the two $\chi_c$ 
mesons are produced back-to-back at the lowest order. 
This is not so in the $k_T$-factorization approach discussed here.
In Fig.\ref{fig:dsig_dptsum} we show distributions of the transverse momentum
of the meson pair, $p_{T,sum}$. 
The distribution for the $\chi_c(1) \chi_c(1)$ extends to large pair
transverse momenta, which is related to the corresponding vertex structure.

\begin{figure}[!h]
\includegraphics[width=8cm]{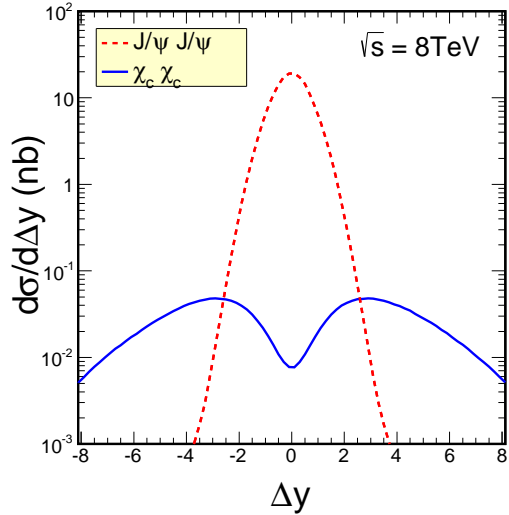}
\caption{Distributions in the rapidity difference 
between two $J/\psi$ (dashed line) and for the sum 
over all $\chi_c \chi_c$ combinations multiplied by 
combined branching fractions.}
\label{fig:dsig_dydiff_jpsijpsi}
\end{figure}

The $\chi_c$ mesons radiatively decay into $J/\psi$ mesons.
The double feed down leads to a new contribution to the $J/\psi J/\psi$
channel. The direct $J/\psi J/\psi$ contribution is more than order
of magnitude larger than the feed-down contribution. 
However, the $\chi_c \chi_c$ contribution has its own specificity. 
In Fig.\ref{fig:dsig_dydiff_jpsijpsi} we show distribution in rapidity 
difference for all $\chi_c \chi_c$ contributions weighted 
by branching fractions into $J/\psi$ channel (solid line)
compared to the standard direct $J/\psi J/\psi$ contribution (dashed
line). At large rapidity difference the feed-down contribution dominates
over the contribution of the standard mechanism. 
Here we assumed that the $J/\psi$'s from the decay will be collinear to
their parent $\chi_c$'s. How important is the feed-down contribution for 
different experimental situations will be discussed elsewhere \cite{paperJPsi}.

\section{Conclusions}

We have made a first exploratory study of $\chi_c$ pair production
in proton-proton collisions.
The $g^* g^* \to \chi_c(J_i) \chi_c(J_j)$ amplitudes for off-shell gluons
and different spin combinations $J_i, J_j = 0,1,2$ are calculated based
on $g^* g^* \to \chi_c(J)$ verticies calculated within the color-singlet 
nonrelativistic pQCD approach. In this approach the vertices are proportional to
the derivative of the spatial wave function at the origin $|R'(0)|$. 
The value of this quantity can be obtained from models of the quarkonia
states. Here it has been obtained from the $\chi_c(0) \to \gamma \gamma$ 
branching fraction which was measured experimentally.

We have performed calculations within the $k_T$-factorization approach 
for the $p p \to \chi_c \chi_c X$ process at $\sqrt{s} = 8 \, \rm{TeV}$ using  
Kimber-Martin-Ryskin \cite{Kimber:2001sc} 
type unintegrated gluon distribution based on the MSTW2008 
\cite{MSTW08} collinear gluons.

We have found that the cross sections for different combinations
of $\chi_c$ quarkonia are of a similar size.
The integrated cross sections for different channels are of the order 
of a few nb.
This is of the same order of magnitude as the cross section for 
$J/\psi$ pair production.
This means that a feedown from the double $\chi_c$ decays
$\chi_c \to J/\psi \gamma$ leads to extra nonnegligible contribution
which has to be included in the total prompt production of two $J/\psi$
mesons. Due to specific branching fractions the $\chi_c(1) \chi_c(1)$,
$\chi_c(1) \chi_c(2)$ and $\chi_c(2) \chi_c(2)$ channels are
the dominant ones. The other three contributions can be safly neglected. 

The $\chi_c \chi_c$ contribution to the $J/\psi J/\psi$ final state 
is interesting but goes beyond the scope of the present analysis and 
will be studied in detail in future dedicated analyses.

The salient feature of the $t$ and $u$-channel gluon exchange mechanism 
are the broad distributions in rapidity difference $\Delta y$ between $\chi_c$ mesons.
This is to be contrasted with the narrow $\Delta y$ distribution of $J/\psi$ pairs
at leading order. 
A feed-down from double $\chi_c$ production to the double $J/\psi$
channel is therefore expected to be important at large $\Delta y$ and may
mimic the kinematical behaviour of double parton scattering mechanisms.

\vspace{1cm}

{\bf Acknowledgments}

We would like to thank Sergey Baranov for discussions and comments on the 
manuscript.
This study was partially supported by the Polish National Science Center
grant DEC-2014/15/B/ST2/02528 and by the Center for Innovation and
Transfer of Natural Sciences and Engineering Knowledge in
Rzesz{\'o}w.


\end{document}